\documentclass[prm,longbibliography,twocolumn,amsmath,amssymb,superscriptaddress,floatfix]{revtex4-1}
\usepackage[USenglish]{babel}
\usepackage{blindtext}
\usepackage[utf8]{inputenc}
\usepackage{graphicx}
\usepackage{array} 
\usepackage{bm}
\usepackage{latexsym}
\usepackage{physics}
\usepackage{amsfonts}
\usepackage{bbold}
\usepackage{subfig}
\usepackage{scalerel}
\usepackage{appendix}
\usepackage{ragged2e}

\usepackage[colorlinks = true,
            linkcolor = blue,
            urlcolor  = blue,
            citecolor = blue,
            anchorcolor = blue]{hyperref}
\usepackage{fixltx2e}
\usepackage[usenames,dvipsnames]{color}
\makeatletter
\newcommand*{\rom}[1]{\expandafter\@slowromancap\romannumeral #1@}
\newcommand{\vebm}[1]{
\ifcat\noexpand#1\relax
    {\bm #1}
\else
    {\bf #1}
\fi
}
\makeatother
\begin{document}

\title{Weak $\mathbb{Z}_2$ Supertopology}

\date{\today}

\author{Kirill Parshukov}
\email{k.parshukov@fkf.mpg.de}
\affiliation{Max Planck Institute for Solid State Research, Heisenbergstrasse 1, D-70569 Stuttgart, Germany}

\author{Moritz M. Hirschmann}
\affiliation{
RIKEN Center for Emergent Matter Science, Wako, Saitama 351-0198, Japan}

\author{Andreas P. Schnyder}
\affiliation{Max Planck Institute for Solid State Research, Heisenbergstrasse 1, D-70569 Stuttgart, Germany}

\begin{abstract}
Crystal symmetries can enforce all bands of a material to be topological, a property that is commonly referred to as ``supertopology". Here, we determine the symmetry-enforced $\mathbb{Z}_2$ supertopologies of non-magnetic centrosymmetric materials with weak and strong spin-orbit coupling (SOC). For weak (i.e., negligible) SOC, crystal symmetries can enforce Dirac nodal lines protected by a $\pi$-Berry phase, while for strong SOC, crystal symmetries can give rise to nontrival weak $\mathbb{Z}_2$ topologies in 2D subplanes of the 3D Brillouin zone. We catalogue all centrosymmetric space groups whose symmetries enforce these  $\mathbb{Z}_2$ supertopologies. Suitable material realizations are identified and experimental signatures of the supertopologies, such as quantum spin Hall states, are being discussed. 
\end{abstract}

\maketitle
\address{}

\section{Introduction}

In the last decade and a half, the study of materials with non-trivial band topology has become an important field of physics~\cite{wieder_nat_rev_22,hasan_kane_review_2010,chiu_review_RMP_2016}. 
The exotic properties of these materials, such as chiral band crossings, anomalous surface states, and topological transport, are described by topological invariants, e.g., Chern numbers or
$\mathbb{Z}_2$
indices~\cite{kane_mele_Z2_PRL_05}.
In order to understand the origin of these topological band features, it is necessary to investigate the connection between  symmetry and topology: 
The crystal symmetries do not only lead to symmetry-related copies of the topological band features, but
 they enforce by themselves certain nontrivial band topologies. For example, screw rotations
 enforce Weyl points on rotation axes~\cite{PhysRevMaterials.5.054202,PhysRevMaterials.5.124202,PhysRevResearch_043165_03}, while glide mirrors 
 enforce nodal lines~\cite{PRM_124204, fang2016topological, bian2016drumhead, xie2021kramers, takahashi2017spinless}.
This property is referred to as ``symmetry-enforced topology"~\cite{watanabe2016filling, takahashi2017spinless, zhao2016nonsymmorphic, Wieder2017FillingEnforced, PRM_124204, PhysRevMaterials.5.054202,PhysRevMaterials.5.124202, wilde2021symmetry, Micklitz2017Sym-Enforced, ono2021z2enriched} and guarantees that bands must be topological irrespective of the chemical composition and other details of the material. 
The principle of symmetry-enforced topology can be used to design or search for new materials with the desired topological characteristics~\cite{PhysRevMaterials.5.124202}.
Another intriguing class of materials is called ``supertopological materials"~\cite{vergniory_supertopology_science_22}, for which all connected bands isolated in energy have a stable topological invariant. Similar property was studied in electric~\cite{jiang2023projective} and superconducting~\cite{PhysRevResearch.3.013288, PhysRevResearch.4.013038, PhysRevX.14.041041} circuits, where all bands carry a non-trivial invariant.
To obtain topological
responses in supertopological materials, a precise tuning of the electronic filling is not necessary,
since all bands are topologically nontrivial. 

So far, symmetry-enforced band topologies due to screw and glide symmetries have been enumerated~\cite{PhysRevMaterials.5.054202,PhysRevMaterials.5.124202} and 
their origin has been explained in terms of three fundamental theorems~\cite{PhysRevResearch_043165_03}. Systematic catalogues of topological materials have revealed a large number of supertopological compounds, where every band has a nontrivial $\mathbb{Z}_2$ Kane-Mele index~\cite{vergniory_supertopology_science_22}. However, these surveys do not explain the underlying mechanism for the supertopology and, moreover, weak $\mathbb{Z}_2$ invariants are not  considered for partial fillings of connected sets of bands.

In this paper, we derive the symmetry-enforced  $\mathbb{Z}_2$ supertopologies in nonmagnetic centrosymmetric materials with weak (i.e., negligible) and strong spin-orbit coupling (SOC).
We find that for strong SOC, crystal symmetries can enforce weak $\mathbb{Z}_2$ topologies in 2D subplanes of the 3D Brillouin zone (BZ), while for weak SOC crystal symmetries
can guarantee the existence of Dirac nodal lines. 
We enumerate all centrosymmetric space groups (SGs), whose symmetries enforce these weak $\mathbb{Z}_2$ topologies and Dirac nodal lines, see Tables~\ref{table:enforced_weakZ2_2bands} and \ref{table:enforced_weakZ2_4bands}. 

Our approach makes use of the Fu-Kane $\mathbb{Z}_2$ symmetry indicator for centrosymmetric systems~\cite{PhysRevB.76.045302}, which expresses the Kane-Mele $\mathbb{Z}_2$ index~\cite{kane_mele_Z2_PRL_05} in terms of inversion eigenvalues at the time-reversal invariant momenta (TRIMs).
This index can be defined for both strong and weak SOC~\cite{kane_mele_Z2_PRL_05,PhysRevLett.115.036806}, indicating $\mathbb{Z}_2$ topology of a (partial) band gap or the existence
of Dirac nodal lines, respectively.  
Centrosymmetric non-magnetic materials exhibit a space-time inversion symmetry $P\mathcal{T}$, which for  strong (weak) SOC squares to $-1$ ($+1$). As a consequence,
the electronic bands are doubly degenerate, either due to Kramers theorem (when $(P\mathcal{T})^2 = -1$)  or due to spin degeneracy (when $(P\mathcal{T})^2 = +1$). 
Hence, at the TRIMs the $\mathbb{Z}_2$ indicator is only well-defined for even fillings. In addition, depending on the considered space group, extra crystal symmetries
can lead to fourfold degeneracies at the TRIMs, such that the $\mathbb{Z}_2$ indicator is only defined for fillings that are multiples of four.
In our analysis we need to treat these two cases separately, see Secs.~\ref{sec_two_fold} and~\ref{sec_four_fold}, respectively. 

The remainder of this paper is organized as follows.
In Sec.~\ref{sec_Z2_indicator} we review the definition and significance of the $\mathbb{Z}_2$ indicator both for strong and weak SOC.
The space groups whose crystal symmetries enforce nontrivial $\mathbb{Z}_2$ topologies
are determined in Secs.~\ref{sec_two_fold} and~\ref{sec_four_fold}, see Tables~\ref{table:enforced_weakZ2_2bands} and \ref{table:enforced_weakZ2_4bands}.
In both of these sections, we focus first on a particular example, i.e., space groups 15 (C2/c) and 135 (P4$_2$/mbc), respectively, and then generalize the deliberations to other space groups.
Material examples are discussed throughout all of these three sections and summarized in Table~\ref{table:Materials}. 
We conclude in Sec.~\ref{sec_conclusions} with some final remarks and directions for future research.
Some technical details are given in the appendices.

\section{$\mathbb{Z}_2$ symmetry indicator}
\label{sec_Z2_indicator}

For inversion-symmetric materials with time-reversal symmetry ($\mathcal{T}^2=\pm 1$) one can define a $\mathbb{Z}_2$ indicator in terms of inversion eigenvalues at the TRIMs~\cite{PhysRevB.76.045302,PhysRevLett.115.036806,PhysRevLett.98.106803,Balents_Moore_PRB_07,Turner2012QuantizedResponse}.
This works for any filling,
for which there is a full band gap
at all considered TRIMs.
For a given 2D plane containing the four TRIMs $\Gamma_i$ ($i \in \{a,b,c,d\})$ the $\mathbb{Z}_2$ indicator
is defined as
\begin{equation}
    (-1)^{\nu} = \prod_{i \in \{a,b,c,d\}} \delta_i, \ \ \delta_i = \prod_{m=1}^{N/2} \xi_{2m}(\Gamma_i),
    \label{Z2_invariant}
\end{equation}
where $\delta_i$ is a product of inversion eigenvalues $\xi_{2m}$ at the TRIM $\Gamma_i$ of the doubly degenerate bands $2m$, and $N$ is the number of occupied bands. In the following, we assume $N$ to be even to ensure an insulating gap at the TRIMs.

In the case of negligible SOC with $(P\mathcal{T})^2=+1$,  
the indictor $\nu$, Eq.~\eqref{Z2_invariant}, describes the value of  the quantized Berry phase~\cite{Taylor2011InversionSymTI, Chen2012BulkTopological, Alexandrinata2014WilsonLoop, Bouhon2017, Bouhon2021, PhysRevResearch_043165_03}, evaluated along any time-reversal symmetric loop $C$ passing through the four TRIMs \cite{PhysRevLett.115.036806}
\begin{equation}
    C = c_{ab} - \overline{c_{ab}} - (c_{cd} - \overline{c_{cd}}),
    \label{Berry_phase_contour}
\end{equation}
where $c_{ab}$ and $\overline{c_{ab}}$ are time-reversed paths between the two TRIMs $\Gamma_a$ and $\Gamma_b$. Specifically, the Berry phase $\gamma \in \{ 0, \pi \}$ is given by (see Ref.~\onlinecite{PhysRevLett.115.036806} and Appendix~\ref{BerryPhaseParity})
\begin{equation}
    e^{i\gamma} = (-1)^{\nu}.
    \label{Z2_BerryPhase}
\end{equation}
The non-trivial value protects a nodal line piercing a surface enclosed by the contour. In other words, the contour and line degeneracy are linked. The degeneracy is protected by the $(P\mathcal{T})^2=+1$ symmetry.

SOC splits spin degeneracies in general. For time-reversal symmetric systems, it implies $\mathcal{T}^2=-1$. Therefore, the symmetry $(P\mathcal{T})^2=+1$ changes to $(P\mathcal{T})^2=-1$, and
the line degeneracy discussed above is gapped out. At the same time, all the electronic bands stay doubly-degenerate due to the Kramers theorem. 
If the indicator [Eq.~\eqref{Z2_invariant}] is defined for four TRIMs in a plane and the SOC opens a gap, the value determines the Kane-Mele $\mathbb{Z}_2$ invariant.
The topological phase implies the emergence of helical states at the material surface.

\section{Twofold degeneracies at TRIMs}
\label{sec_two_fold}

Crystal symmetries can enforce degeneracies at high-symmetry points in the Brillouin zone~\cite{PhysRevMaterials.5.054202,PhysRevMaterials.5.124202}. To begin with we consider the space groups with four two-fold degenerate TRIMs due to the $P\mathcal{T}$ symmetry. Inversion eigenvalues for degenerate states are the same and cannot be determined by the symmetries. However, the $\mathbb{Z}_2$ indicator can be enforced if the eigenvalues are related between different TRIMs~\cite{masters_thesis}.

\subsection{Space Group 15 (C2/c)}

We consider space group 15 as an example of a space group enforcing a $\mathbb{Z}_2$ invariant to be non-trivial for $4\mathbb{N}+2$ occupation. 

The monoclinic space group 15 (C2/c) is generated by translations along $x$-, $y$-, $z$-axes $t(1, 0, 0)$, $t(0, 1, 0)$, $t(0, 0, 1)$ respectively, glide mirror symmetry $M_{010}(0, 0, \tfrac{1}{2})$ along the y-axis, inversion $P$, and the translation $t(\tfrac{1}{2}, \tfrac{1}{2}, 0)$, which is characteristic for the base-centered lattice~\cite{AroyoPerezMatoCapillasKroumovaIvantchevMadariagaKirovWondratschek+2006+15+27}. 
The action of symmetries can be represented as follows
\begin{align}
    M_{010}(0, 0, \tfrac{1}{2}): (x, y, z) &\to (x, -y, z+\tfrac{1}{2}), \\
    P: (x, y, z) &\to (-x, -y, -z),
    \label{SG15_symmetries}
\end{align}
where we consider the spinless representations.

\begin{figure}
  \centering
  \includegraphics[width=0.48\textwidth]{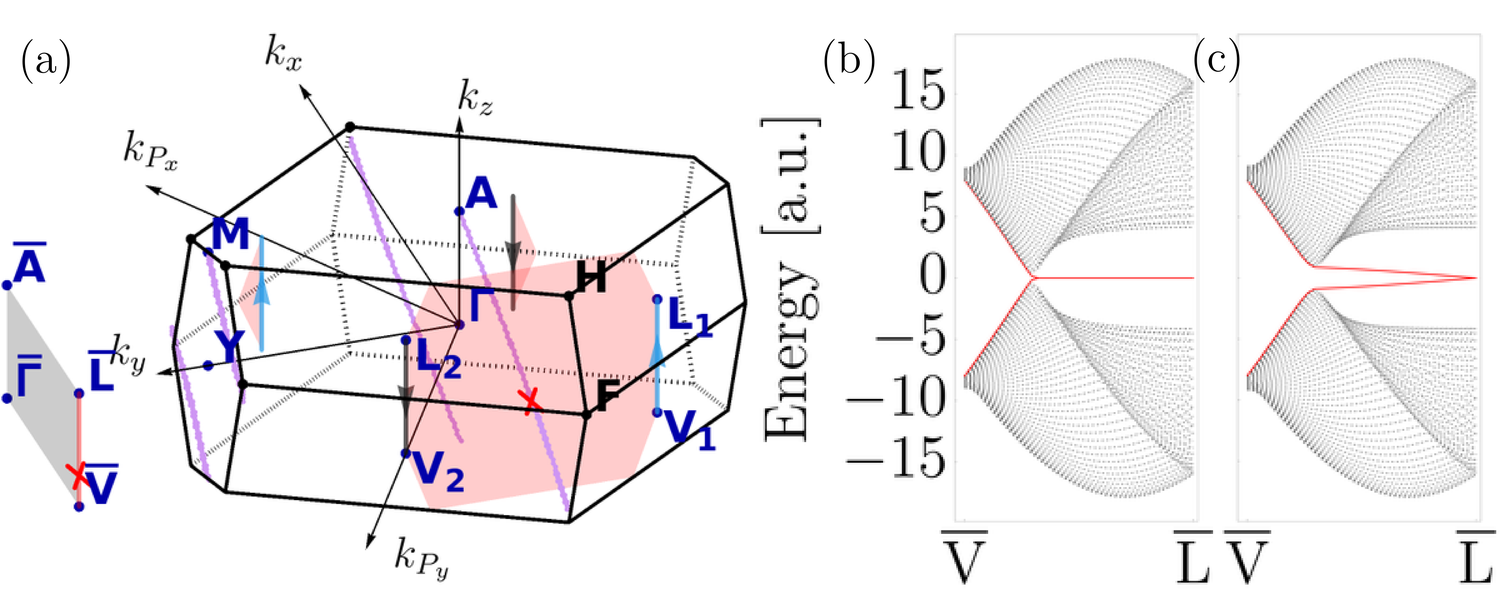}
  \caption{\justifying 
  (a) Space group 15 BZ. TRIMs are shown in blue. Axes $k_x,k_y,k_z$ are conventional, whereas $k_{Px},k_{Py},k_z$ are primitive. The plane L$_1$L$_2$V$_2$V$_1$ is projected onto $\overline{\text{L}}\overline{\text{V}}$ line under the introduction of the open boundary in $y$ direction. The symmetry-enforced nodal line is shown in purple. Blue and black arrows enclose the surface with $\pi$ Berry phase. (b), (c) Bands of a symmetric TB model without and with SOC (see Appendix~\ref{TBSG15}). The SOC opens the gap inside the plane. Red bands are the surface states.
}
    \label{BZ_SG15_plus_Slab}
\end{figure}

A sketch of the BZ is shown in Fig.~\ref{BZ_SG15_plus_Slab} (a) following Bilbao crystallographic server notation~\cite{AroyoPerezMatoCapillasKroumovaIvantchevMadariagaKirovWondratschek+2006+15+27,HINUMA2017140,SETYAWAN2010299}. For more information about the basis vectors, see Appendix~\ref{Appendix_SG15}. We are interested in a plane in the BZ with four TRIMs $\text{L}_1(-\pi, -\pi, \pi)$, $\text{L}_2(-\pi, \pi, \pi)$, $\text{V}_1(-\pi, -\pi, 0)$ and $\text{V}_2(-\pi, \pi, 0)$.

The little groups for all of L and V points contain only inversion symmetry. Generally, it means that the little group is abelian with 1D irreducible representations. 
It ensures the absence of any symmetry-protected degeneracies.

The mirror symmetry $M_{010}(0, 0, \tfrac{1}{2})$ acts on a vector $\mathbf{k}=(k_x, k_y, k_z)$ in the BZ relating it with $R_{M_{010}}\mathbf{k}=(k_x, -k_y, k_z) = \tilde{\mathbf{k}}$, where $R_{M_{010}}$ is a transformation matrix in $k$-space. Therefore, the mirror reflection relates $V_1$ and $V_2$ points, $L_1$ and $L_2$ points.

Now we consider a Bloch state $|\psi(\mathbf{k})\rangle$ at momentum $\mathbf{k}$. It is related by mirror reflection symmetry with the state $D_{M_{010}}|\psi(\mathbf{k})\rangle = |\tilde{\psi}(R_{M_{010}}\mathbf{k})\rangle$, where $D_{M_{010}}$ is symmetry representation in Hilbert space. These two states have the same energy.

At TRIMs parities $\xi$ of Bloch states $|\psi(\Gamma_i)\rangle$ are well-defined, because $\Gamma_i = -\Gamma_i + G$, where $G$ is a reciprocal vector. Importantly, the parities at $|\psi(\mathbf{k})\rangle \ \text{and} \ D_{M_{010}}|\psi(\mathbf{k})\rangle$ are connected. The connection can be deduced from the commutation relation $D_P D_{M_{010}} = D_{M_{010}} D_P D_{t(0, 0, 1)}$, where $D_P, D_{t(0, 0, 1)}$ are representations of inversion and translation operators, respectively. For proof of the statement, see Appendix \ref{Appendix_SG15_parities}. When applied to a Bloch state at $\mathbf{k}$ the expression is written as
\begin{equation}
    D_P D_{M_{010}} |\psi(\mathbf{k})\rangle = e^{i k_z} D_{M_{010}} D_P |\psi(\mathbf{k})\rangle.
    \label{SG15_commutation}
\end{equation}
Now if the parity of the state $|\psi(\mathbf{k})\rangle$ is defined at $\mathbf{k}$ and equal to $\xi$, then parity of $D_{M_{010}}|\psi(\mathbf{k})\rangle$ is $e^{i k_z} \xi$.

Applying the argument to L and V points we conclude that 
\begin{align}
    \xi(\text{L}_1) = -\xi(\text{L}_2), \ \ \xi(\text{V}_1) &= \xi(\text{V}_2), \\ \xi(\text{L}_1)\xi(\text{L}_2)\xi(\text{V}_1)\xi(\text{V}_2) &= -1.
    \label{parity_relations}
\end{align}
The Berry phase Eq.~\eqref{Z2_BerryPhase} for the contour Eq.~\eqref{Berry_phase_contour} passing the four TRIMs is enforced to be $\pi$ for every band. The quantized non-trivial Berry phase implies the presence of the nodal line piercing the surface enclosed by the contour.

A SOC-free material with symmetries of the SG 15 has a nodal line between the conduction and valence bands if an odd number of Kramers partners is occupied $N = 4\mathbb{N} + 2$. 
The presence of the degeneracy is due to the band inversion at one of the TRIMs. The SOC lifts the line degeneracy opening a gap inside the plane with L and V points. The two bands are necessarily topological, and the gap between them is non-trivial. Formally, the relation between inversion eigenvalues Eq.~\eqref{parity_relations} is true in both regimes. For systems with strong SOC, the actions of time-reversal and mirror acquire a spin-part, transforming the spin degree of freedom. However, the inversion is not affected, and the relation Eq.~\eqref{SG15_commutation} is satisfied. 
The Dirac nodal line can be considered as a parent state for the topological insulating phase.
Every disconnected band in the plane with L, V points has the non-trivial $\mathbb{Z}_2$ invariant. Therefore, every second gap in the plane ($4\mathbb{N} + 2$ occupation) has a gapless surface state.

We constructed a tight-binding (TB) model with symmetries of the space group 15 with and without SOC Fig.~\ref{BZ_SG15_plus_Slab} (b)-(c) [see Appendix~\ref{TBSG15}]. There is an enforced crossing inside the plane, corresponding to the piercing nodal line. The degeneracy is split under SOC opening a gap inside the L-V plane. 
For the constructed TB models, we introduced a termination direction along the $y$-axis to observe surface states due to non-trivial topology, see Fig.~\ref{BZ_SG15_plus_Slab} (b)-(c). 

As TRIMs are defined by $\Gamma_i = -\Gamma_i + G$, there are several options to choose a plane (contour) with four discussed points and non-trivial $\mathbb{Z}_2$ invariant. For SG 15 we can consider $\text{L}_1(\pi, \pi, -\pi)$ and $\text{V}_1(\pi, \pi, 0)$ instead, then the contour with $\pi$ Berry phase encloses a nodal line pinned to high-symmetry line AM.

\subsection{Generalizations to other space groups}
To generalize the approach, we list criteria that we used to find all centrosymmetric space groups, whose symmetry enforces a non-trivial invariant in a 2D submanifold of the 3D BZ.
\begin{itemize}
    \item The $\delta_i$ product should be well-defined at four TRIMs [see Eq. \eqref{Z2_invariant}]. For the $4\mathbb{N}+2$ occupation it is possible only if there is no enforced degeneracy at the TRIMs, except the spin-degeneracy. It means there are only one dimensional irreps of the TRIMs single-valued little groups in case of the negligible SOC, and all double-valued group irreps must be two-dimensional for strong SOC. 
    \item The space group must contain a symmetry element, that relates the TRIMs. Moreover, any TRIM among the four should be related to at least one another. This allows us to relate inversion eigenvalues of electron states at the TRIMs. Importantly, these states also have the same energy due to the symmetry. It is common in the literature to give the same label to such high-symmetry points.
    \item The space group must be non-symmorphic. Only in this case, it is possible to pair states with opposite parities at the symmetry-related TRIMs [see Eq. \eqref{SG15_commutation}]. This condition is necessary to obtain a non-trivial value of the invariant Eq. \eqref{Z2_invariant}.
    \item In the case of strong SOC the $\mathbb{Z}_2$ invariant is defined for a gapped phase, therefore the space group must not enforce any additional degeneracies inside the plane with non-trivial invariant.
\end{itemize}
We found eight space groups with a non-trivial invariant in a 2D submanifold without SOC, see Table~\ref{table:enforced_weakZ2_2bands}. Three of them SG 70, 88, and 141 have four non-equivalent TRIMs related by a four-fold symmetry. Five SGs 15, 66, 74, 84, and 131 have two pairs of TRIMs related by two-fold symmetries.
Moreover, two space groups enforce additional degeneracy at the TRIMs, which is not compatible with the first criterion. However, when  SOC is strong, the degeneracy is lifted.
\begin{table}[ht]
\centering
\renewcommand{\arraystretch}{1.2} 
\begin{tabular}{||p{1.2cm} | p{2cm}| p{2.4cm}|p{1cm}|p{1cm}||} 
 \hline
 SG & $\prod\limits_i\delta_i=1$ & $\prod\limits_i\delta_i=-1$ & SOC & No SOC\\ [0.5ex] 
 \hline\hline
 15 & (V$_1$, V$_2$) & (L$_1$, L$_2$) & \checkmark & \checkmark\\ 
 \hline
 66 & (S$_1$, S$_2$) & (R$_1$, R$_2$) & \checkmark & \checkmark\\ 
 \hline
 70 & &(L$_1$, L$_2$, L,$_3$ L$_4$) & \checkmark & \checkmark \\
 \hline
  74 & (R$_1$, R$_2$) & (S$_1$, S$_2$) & \checkmark & \checkmark\\ 
 \hline
 84 & (X$_1$, X$_2$) & (R$_1$, R$_2$)  & \checkmark & \checkmark \\ 
 \hline
  88 & &(N$_1$, N$_2$, N$_3$, N$_4$) & \checkmark & \checkmark \\
 \hline
 131 & (X$_1$, X$_2$) & (R$_1$, R$_2$) & \checkmark & \checkmark  \\ 
 \hline
  141 & &(N$_1$, N$_2$, N$_3$, N$_4$)  & \checkmark & \checkmark \\
 \hline
 203 & &(\textbf{L}$_1$, \textbf{L}$_2$, \textbf{L}$_3$, \textbf{L}$_4$)  & \checkmark &  \\
 \hline
  227 & & (\textbf{L}$_1$, \textbf{L}$_2$, \textbf{L}$_3$, \textbf{L}$_4$) & \checkmark & \\
 \hline
\end{tabular}
\caption{\justifying Symmetry-enforced $\mathbb{Z}_2$ at $4\mathbb{N}+2$ filling. TRIMs highlighted in bold can be 4-fold degenerate for negligible SOC. The degeneracy is lifted by the strong SOC.}
\label{table:enforced_weakZ2_2bands}
\end{table}
\section{Fourfold degeneracies at TRIMs}
\label{sec_four_fold}

If crystal symmetries enforce four-fold degeneracies at TRIMs and relate the momenta, the
$\delta$ values [Eq. \eqref{Z2_invariant}] must be the same at two related points. Therefore, to enforce the invariant to be non-trivial, symmetries of a TRIM's little group must relate inversion eigenvalues of degenerate states. In Refs.~\cite{PhysRevMaterials.5.054202,PhysRevMaterials.5.124202} the authors developed this approach for two possible cases of finite weak and strong SOC. For the strong SOC all considered TRIMs must have a four-fold degeneracy enforced by crystal symmetries. If the SOC is weak the degeneracies can be two-fold, however, in the absence of SOC the bands must be four-fold degenerate. The latter case is unstable, as strong SOC might exchange bands. 
In this work, we complete the space group catalog.

\subsection{Space Group 135 (P4$_2$/mbc)}

We discuss the approach considering the space group 135. The spinless representations of glide mirror symmetries are given by
\begin{align}
    M_{100}(\tfrac{1}{2},\tfrac{1}{2},0)&:\ (-x +\tfrac{1}{2}, y+\tfrac{1}{2}, z),\\
    M_{010}(\tfrac{1}{2},\tfrac{1}{2},0)&:\  (x +\tfrac{1}{2}, -y+\tfrac{1}{2}, z) ,\\
    M_{110}(\tfrac{1}{2},\tfrac{1}{2},\tfrac{1}{2})&: \ (-y +\tfrac{1}{2}, -x+\tfrac{1}{2}, z+\tfrac{1}{2}).
\end{align}
With this we can write relations between representations acting on Bloch states
\begin{align}
    D_PD_{M_{100}}&=e^{i(-k_x+k_y)}D_{M_{100}}D_P, \\
    D_PD_{M_{010}}&=e^{i(k_x-k_y)}D_{M_{010}}D_P, \\
    D_PD_{M_{110}}&=e^{i(-k_x-k_y+k_z)}D_{M_{110}}D_P, \\
    D_{M_{100}}D_{M_{010}}&=(-1)^{f}e^{i(k_x-k_y)}D_{M_{010}}D_{M_{100}},\\
    D_{M_{010}}^2&=(-1)^{f}e^{i k_x},\\
    D_{\mathcal{T}}^2&=(-1)^{f},
\end{align}
where $f = 1$ for spin-$\tfrac{1}{2}$ particles and $f = 0$ for the spinless case.
The TRIMs R$_1(\pi,0,\pi)$, R$_2(0,\pi,\pi)$, X$_1(\pi,0,0)$ and X$_2(0,\pi,0)$ are invariant under $P, \  M_{100} \ \text{and} \ M_{010}$ symmetries. For these points the relation $\{D_P,D_{M_{100}}\}=0$ enforces the degeneracy with $(+-)$ inversion eigenvalues. In both cases of strong and negligible SOC the degeneracy is fourfold, and $\delta = -1$ in Eq. \eqref{Z2_invariant}.

At the A$(\pi, \pi,\pi)$ and M$(\pi,\pi,0)$ points the same symmetries give the relations $[D_P,D_{M_{010}}]=0$, $D_{\mathcal{T}M_{010}}^2 = -1$. For the spinless case one already can conclude the presence of the degeneracy due to the Kramers theorem with the same inversion eigenvalues. In the spinful case we additionally can write $[D_P,D_{M_{100}}]=\{D_{M_{100}},D_{M_{010}}\}=0$. Together the relations enforce four orthogonal states with the same inversion eigenvalue at the same energy $\psi, \  P\mathcal{T}\psi, \ M_{010} \psi, \ \mathcal{T}M_{010}\psi$. For the M point it implies $\delta = 1$. 

At the A$(\pi, \pi,\pi)$ and Z$(0,0,\pi)$ points the symmetry $M_{110}$ enforces degeneracy with opposite inversion eigenvalues due to the relation $\{D_P,D_{M_{110}}\}=0$. $\delta_Z = -1$, however, at A point there is already fourfold degeneracy with the same inversion eigenvalues, therefore, the total degeneracy at the A point is eight.

\begin{figure}
    \includegraphics[width=0.48\textwidth]{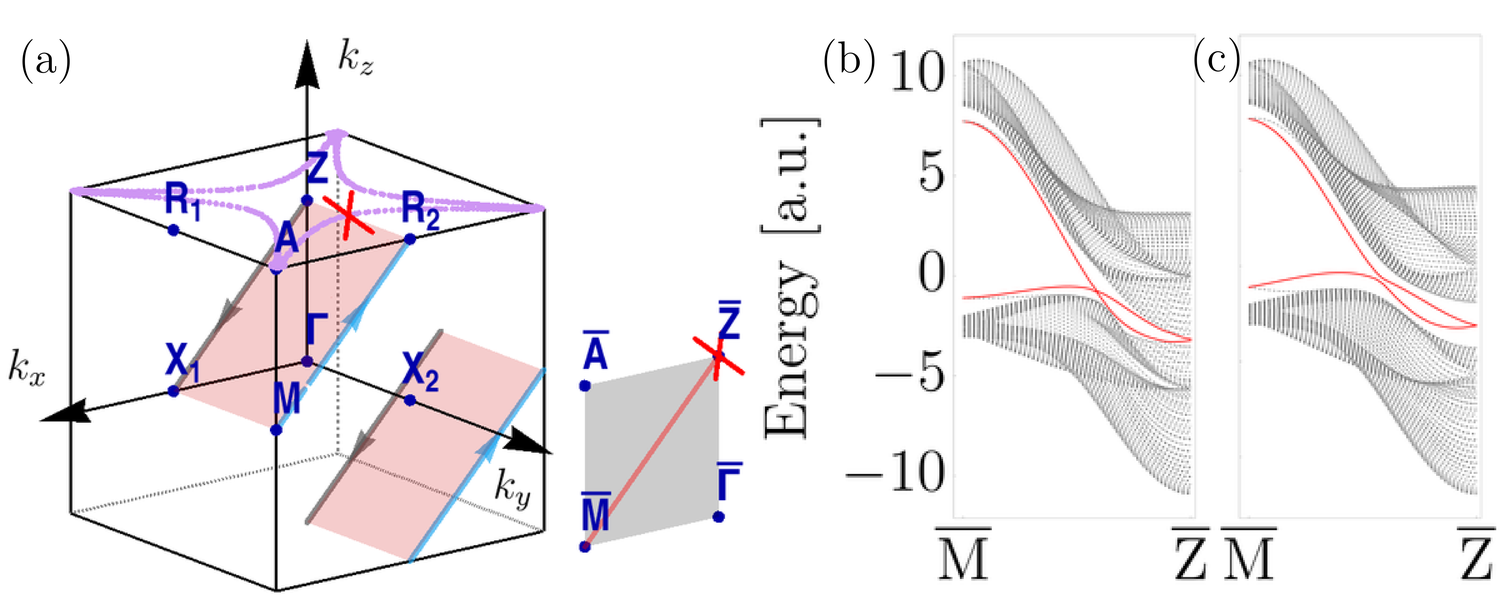}
  \caption{\justifying (a) Space group 135  BZ. The non-trivial plane is shown in red is projected onto 2D surface BZ. The symmetry-enforced nodal line is shown in purple, it crosses the plane along the ZR$_2$ line. (b), (c) Bands of a symmetric TB model without and with SOC (see Appendix~\ref{TBSG135}). The SOC opens the gap inside the plane. Red bands are the surface states.}
    \label{BZ_SG135_plus_Slab}
\end{figure}

Bands in the plane R-Z-X-M host the non-trivial $\mathbb{Z}_2$ invariant. In case of the negligible SOC the plane is pierced by a nodal line, pinned to the $k_z=\pi$ plane. For a TB model (see Appendix~\ref{TBSG135}) the location of the line in BZ is shown in Fig.~\ref{BZ_SG135_plus_Slab} (a). In a model with termination in $y$ direction, the bulk bands are projected onto $k_x-k_z$ plane. The projected nodal line crosses $\overline{\text{MZ}}$ at $\overline{\text{Z}}$. The SOC opens the gap at the point and the helical surface states emerge [Fig.~\ref{BZ_SG135_plus_Slab} (b-c)].
The discussed results are summarized in Table~\ref{table:enforced_weakZ2_4bands}.

\subsection{Generalizations to other space groups}
Four-fold degeneracies at TRIMs can be enforced by the non-symmorphic symmetries. The symmetries relate inversion eigenvalues between degenerate states. We look for the centrosymmetric space groups that have TRIMs with the non-symmorphic little groups. We say that a 2D surface with four 4-fold degenerate TRIMs without SOC is topological, if the $\mathbb{Z}_2$ invariant and the corresponding Berry phase are non-trivial for the $8\mathbb{N}+4$ occupation.
If the surface is the plane we also consider the effect of the SOC.
There are two possibilities, as the degeneracies can persist or be lifted. In the former case the $\mathbb{Z}_2$ Kane-Mele invariant is enforced only by the symmetries and does not depend on a specific material. An example will be the SG 135 with four TRIMs in the plane R-Z-X-M. In the latter case, the invariant is enforced for a small, but finite SOC. The coupling can lift the degeneracies, but should not invert bands. The argument becomes weaker here. For instance, in the SG 60, the four-fold degeneracy at the S point is lifted under the SOC. 

Moreover, symmetries can enforce degeneracies larger than four. If it happens without SOC, the system with the interaction can still have four TRIMs in the plane with four-fold degeneracies. In SG 61, the eight-fold degeneracy at the R point is lifted to four two-fold degenerate bands if the SOC is present.

In all examples of space groups with the enforced $\mathbb{Z}_2$ invariant that we found, see Table~\ref{table:enforced_weakZ2_4bands}, there is one TRIM, where inversion commutes with a symmetry that enforces degeneracy. At the other three TRIMs, the inversion anticommutes with the symmetry operator, enforcing degeneracy with opposite inversion eigenvalues.

\begin{table}
\centering
\renewcommand{\arraystretch}{1.2} 
\begin{tabular}{||p{1cm} | p{1cm}| p{3cm}|p{1cm}|p{1cm}||} 
\hline
 SG & $\delta_i=1$ & $\delta_i=-1$& SOC \cite{PhysRevMaterials.5.054202,PhysRevMaterials.5.124202} & No SOC\\ [0.5ex] 
 \hline\hline
 52 & \textit{T} & X, Y, Z, \textbf{S}, U, R &  & \checkmark \\ 
 \hline
 56 & R & X, Y, Z, S, \textbf{U, T} & \checkmark & \checkmark  \\ 
 \hline
  60$^*$ & \textit{S} & X, Y, Z, \textbf{T, U, R} & & \checkmark \\ 
 \hline
 61 & \textbf{R} & X, Y, Z, \textbf{T, U, S}& \checkmark & \\ 
 \hline
  62 & U & X, Y, Z, T, \textbf{R, S}  & \checkmark & \checkmark\\ 
 \hline
  135 & M & X, R, Z & \checkmark & \checkmark \\
 \hline
  138 & A & X, \textbf{R}, Z, M & \checkmark & \checkmark\\ 
 \hline
\end{tabular}
\caption{\justifying  Symmetry-enforced $\mathbb{Z}_2$ at $8\mathbb{N}+4$ filling. TRIMs highlighted in bold are 8-fold degenerate for the negligible SOC. The degeneracy is lifted by the strong SOC. With italic we showed TRIMs with 4-fold degeneracies that are lifted under the SOC interaction. The asterisk $^*$ represents the fact that there is no possible loop passing X, Y, Z, S points with symmetry-enforced $\pi$ Berry phase (see Appendix~\ref{BerryPhaseContours}).}
 \label{table:enforced_weakZ2_4bands}
\end{table}

\section{Material examples}
We implemented a material search to look for compounds with symmetry-enforced $\mathbb{Z}_2$ quantum spin Hall states (see Table~\ref{table:Materials} \textcolor{black}{and Appendix~\ref{candidates}}). We considered examples of compounds tabulated in the Inorganic Crystal Structure Database (ICSD)~\cite{gledocs_11858_9746} with a number of different elements less than four and a heavy element with atomic number $Z>54$. In this case, the SOC is large enough to open a gap. We did not consider compounds with f-electrons close to the Fermi level to avoid the effects of electron correlations. As mentioned before, every second gap must have the non-trivial $\mathbb{Z}_2$ invariant, so we restricted our search to materials with the non-trivial gap close to the Fermi level (with only one exception of PbWO$_4$ -- a trivial insulator). Also, the number of bands near the Fermi level should not be large to resolve the non-trivial gap.

\begin{figure}[t]
    \centering
\includegraphics[width=0.48\textwidth]{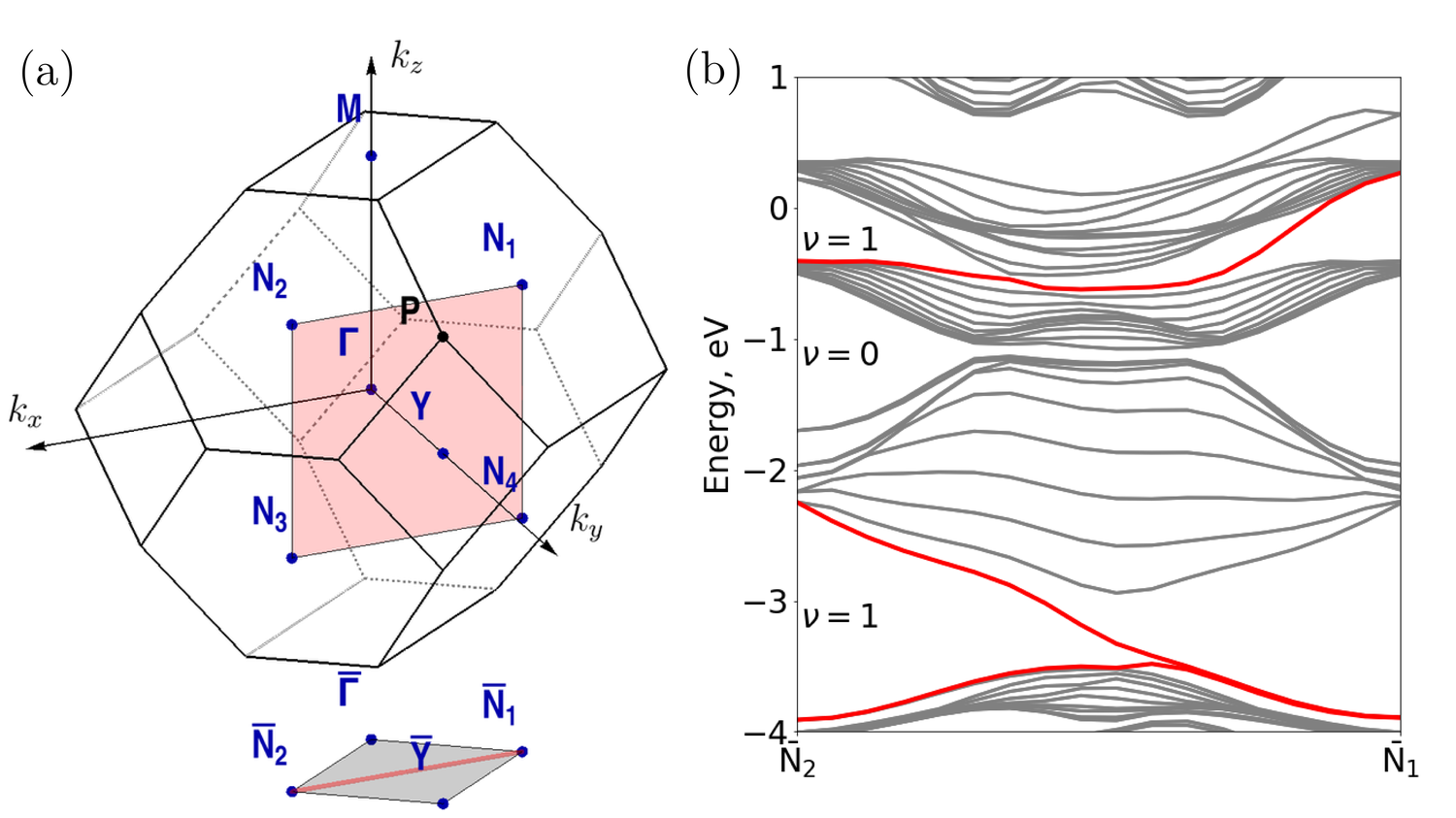}
\caption{\justifying 
(a) Brillouin zone (BZ)
for space group 141.
The non-trivial plane is shown in red, projected onto a surface BZ. (b) TaP slab bands with  10 unit cell layers stacked in the $z$ direction. The surface states in non-trivial bulk gaps are shown in red.}
\label{Slab_PTa}
\end{figure}

The compound TaP in SG 109 (I4$_1$md) is a Weyl semimetal~\cite{PhysRevX.5.011029, doi:10.1126/sciadv.1501092}. However, not much is known about its other phase with the symmetries of SG 141 (I4$_1$/amd) and the structure type $\beta$-NbP \cite{TaP_paper} (see ICSD 108656 \cite{gledocs_11858_9746}).
Its BZ has four N points N$_1(\pi,0,0)$, N$_2(0,\pi,0)$, N$_3(0,\pi,\pi)$ and N$_4(\pi,0,\pi)$ in the primitive reciprocal basis. 
In a slab geometry with termination in the $z$ direction N$_1$ point projects onto N$_4$, and N$_2$ onto N$_3$. We computed the band structure (Fig.~\ref{Slab_PTa} \textcolor{black}{and Fig.~\ref{TaP_bands}}) with the Quantum ESPRESSO software~\cite{Giannozzi_2009,Giannozzi_2017,TaPpseudo} for a supercell with 10 unit cells stacked in the $z$ direction. In every second bulk gap there are helical surface states corresponding to the symmetry-enforced $\mathbb{Z}_2$ invariant. 

Another representative of materials in SG 141 (I4$_1$/amd) is BW~\cite{Kieling1947TheCS} \textcolor{black}{(see Fig.~\ref{BW_bands})}. It becomes superconducting at $T_c = 4.3~\text{K}$~\cite{KAYHAN20121656} and can be an example of a symmetry-enforced topological superconductor.

We found an intriguing material example PbWO$_4$ used in electromagnetic detectors for high-energy physics research~\cite{BORISEVICH2005101}. It grows in a single crystal form with symmetries of SG 88~(I4$_1$/a)~\cite{Sleight:a09322,KAUROVA201656}. Though it is a trivial insulator, the next gap below the Fermi level is enforced to be topological.
\begin{table}[ht]
\centering
\renewcommand{\arraystretch}{1.2} 
\begin{tabular}{||p{.5cm} p{1.5cm} |p{5.5cm}||} 
 \hline
 SG & & Materials \\ [0.5ex] 
 \hline\hline
 15 & (C2/c) & P$_2$Pt$_5$~\cite{R100000136-I1360861714460395904}, Ca$_5$Pt$_2$~\cite{PALENZONA198149}\\ 
 \hline
 52 & (Pnna) & Sr$_2$Bi$_3$~\cite{MERLO1994149,PhysRevMaterials.5.124202}\\
 \hline
 62 & (Pnma)&Ir$_2$Si~\cite{Schubert1960,PhysRevMaterials.5.124202}\\
 \hline
  74 &(Imma)& CsHg$_2$~\cite{Deiseroth}\\ 
 \hline
  88 &(I4$_1$/a)& PbWO$_4$~\cite{KAUROVA201656}\\
 \hline
  141 &(I4$_1$/amd)&   BW~\cite{KAYHAN20121656}, TaP\cite{TaP_paper} \\
 \hline
  227 &(Fd-3m) &  Au$_2$Na~\cite{ZACHWIEJA1993171}, CaPt$_2$~\cite{Wood:a02272}, BaPt$_2$~\cite{Wood:a02272}, YPt$_2$~\cite{KRIKORIAN1971271}\\
 \hline
\end{tabular}
 \caption{\justifying 
 Materials with symmetry-enforced quantum spin Hall states. \textcolor{black}{For several compounds we computed the band structures and show them in the Appendix~\ref{candidates}. The bands for Sr$_2$Bi$_3$ and Ir$_2$Si are shown in Ref.~\cite{PhysRevMaterials.5.124202}.}
 }
 \label{table:Materials}
\end{table}

\section{Conclusions}
\label{sec_conclusions}

In this work, we investigated the origin of supertopology. We found 17 SGs enforcing topological invariants for every band independently of material characteristics. In several space groups, the symmetry-enforced Dirac nodal line (and its surface state) is the parent state for the quantum spin Hall state if SOC is added. 

In the description of the enforced Kane-Mele invariant, we assumed the presence of the bulk band gap in a plane. This assumption does not demand the bulk gap everywhere in the full BZ. This implies the possible coexistence of the bulk Fermi surface with boundary resonances~\cite{PhysRevB.82.195417}.
If the origin of the Fermi surface is the symmetry-enforced connectivity between the bands~\cite{vergniory_supertopology_science_22}, the bulk gap can be opened with the breaking of the crystalline symmetries. Unless there is a band inversion in the non-trivial plane, the invariant must be preserved and non-trivial. This can be an approach to create symmetry-enforced weak topological insulators.

Approaches used in the paper can be applied to study representation-enforced topology~\cite{masters_thesis}. In this case, one can consider a particular Wyckoff position and assume that bands induced from each of the site symmetry groups do not mix. With these conditions the classification of the space groups can be extended to the classification of the Wyckoff positions that induce bands carrying weak and strong $\mathbb{Z}_2$, and higher-order $\mathbb{Z}_4$ topologies~\cite{PhysRevX.8.031070,doi:10.1126/sciadv.aat0346,doi:10.1126/science.aah6442, masters_thesis}. 

The discussed symmetry indicators can be applied to magnetic systems with $P$ and anti-unitary symmetries. The anti-unitary symmetry (time-reversal with a crystalline operation) must quantize $\mathbb{Z}_2$ topological invariant~\cite{PhysRevB.81.245209}, and the inversion eigenvalues indicate a non-trivial phase. Thus, the material search can be extended to magnetic phases.

With the addition of chiral symmetry, 2D electron systems can host higher-order topology~\cite{PhysRevLett.119.246401}. Therefore, symmetry-enforced higher-order topological phases are possible and require further investigation. 

We study the 2D symmetry indicators as the 3D counterparts can not be enforced due to material dependence of the inversion eigenvalues at $\Gamma$ point. However, in higher dimensional systems, the approach can be applied to enforce 3D symmetry indicators. 

\section*{ACKNOWLEDGMENTS}
We wish to thank Hans Peter Büchler, Kirill Alpin, and Andreas Leonhardt for useful discussions.
K.P. and A.P.S. are funded by the Deutsche Forschungsgemeinschaft (DFG, German Research Foundation) – TRR 360 – 492547816.
M.M.H. is funded by the Deutsche Forschungsgemeinschaft (DFG, German Research Foundation) – project number 518238332.




\appendix

\section{SG 15 -- Conventional and primitive bases}
\label{Appendix_SG15}

\textcolor{black}{In this appendix we describe the transformation
from conventional to primitive bases for SG 15.}
To transform reciprocal vectors from a primitive basis $g_1, g_2, g_3$ to a conventional $k_x, k_y, k_z$ one has to use the matrix 
\begin{align}
C = \begin{pmatrix}
1 &-1 & 0\\
1 & 1 & 0\\
0 & 0 & 1
\end{pmatrix}.
\end{align}
In the main text we consider TRIMs $\text{L}_1^g(\pi, 0, \pi)$, $\text{L}_2^g(0, \pi, -\pi)$, $\text{V}_1^g(\pi, 0, 0)$, and $\text{V}_2^g(0, \pi, 0)$. The superscript $g$ corresponds to the primitive basis $g$. The coordinates of TRIMs in primitive are related to the conventional basis by $\mathbf{k}_{\Gamma_i}$ = $C\mathbf{k}_{\Gamma_i}^g$. 

\section{Parities under two-fold rotation}
\label{Appendix_SG15_parities}
Here we prove the commutation relation Eq. \eqref{SG15_commutation}. Inversion and glide mirror representations are given by
\begin{align}
    (x, y, z) &\xrightarrow{P} (-x, -y, -z) \xrightarrow{M_{010}}(-x, y, -z+\tfrac{1}{2}),\\
    (x, y, z) &\xrightarrow{M_{010}}(x, -y, z+\tfrac{1}{2})
    \xrightarrow{P}(-x, y, -z-\tfrac{1}{2}),
\end{align}
where $M_{010}$ corresponds to $M_{010}(0,0,\tfrac{1}{2})$ glide mirror reflection. 
With these two expressions we conclude 
\begin{align}
    D_P D_{M_{010}} &= D_{t(0, 0, -1)} D_{M_{010}} D_P \\ & =  D_{M_{010}} D_{t(0, 0, -1)} D_P \\ &= D_{M_{010}} D_P D_{t(0, 0, 1)}.
\end{align}
The translation operator $D_{t(0, 0, 1)}$ acts on a Bloch state as a phase multiplication $e^{i k_z}$, and $D_P D_{M_{010}} = e^{i k_z} D_{M_{010}} D_P$.

\section{Berry phase and inversion eigenvalues}\label{BerryPhaseParity}
In this appendix, we follow Ref.~\onlinecite{PhysRevLett.115.036806} to show the relation between inversion eigenvalues of occupied states and the Berry phase. 
We consider two Bloch states related by inversion symmetry $P|u_n(-\mathbf{k})\rangle = e^{i \beta_n(\mathbf{k})}|u_n(\mathbf{k})\rangle$. The Berry connection can be defined as $A(\mathbf{k}) = -i \sum_n \langle u_n(\mathbf{k})|\nabla_{\mathbf{k}}u_n(\mathbf{k})\rangle$, where we sum over occupied bands $n$.
Connection difference at $\mathbf{k}$ and $-\mathbf{k}$ is $A(\mathbf{k})-A(\mathbf{-k}) = -\nabla_{\mathbf{k}} \sum_n \beta_n(\mathbf{k})$.
The Berry phase along a path $c_{ab} - \overline{c_{ab}}$ connecting two TRIMs $\Gamma_a$ and $\Gamma_b$ is an integral of the Berry connection
\begin{equation}
    \gamma_{ab} = -\sum_n [\beta_n(\Gamma_b) - \beta_n(\Gamma_a)] \mod 2\pi.
\end{equation}
The path $\overline{c_{ab}}$ represents the time-reversal partner of $c_{ab}$.

At the same time the product of inversion eigenvalues at a TRIM $\Gamma_i$
\begin{equation}
    \prod_{n} \xi_{n}(\Gamma_i) =\prod_{n}\langle u_n(\Gamma_i)|P| u_n(\Gamma_i)\rangle =
    e^{-i \sum_n \beta_n(\Gamma_i)},
\end{equation}
where $i\in\{a,b\}$.
Therefore, the inversion eigenvalues of occupied states determine the Berry phase along the contour $c_{ab} - \overline{c_{ab}}$
\begin{equation}
    e^{i\gamma_{ab}} = \prod_{n} \xi_{n}(\Gamma_a)\xi_{n}(\Gamma_b).
\end{equation}

\section{Berry phase contours}\label{BerryPhaseContours}

\begin{figure}[ht]
    \centering
    \includegraphics[width=0.48\textwidth]{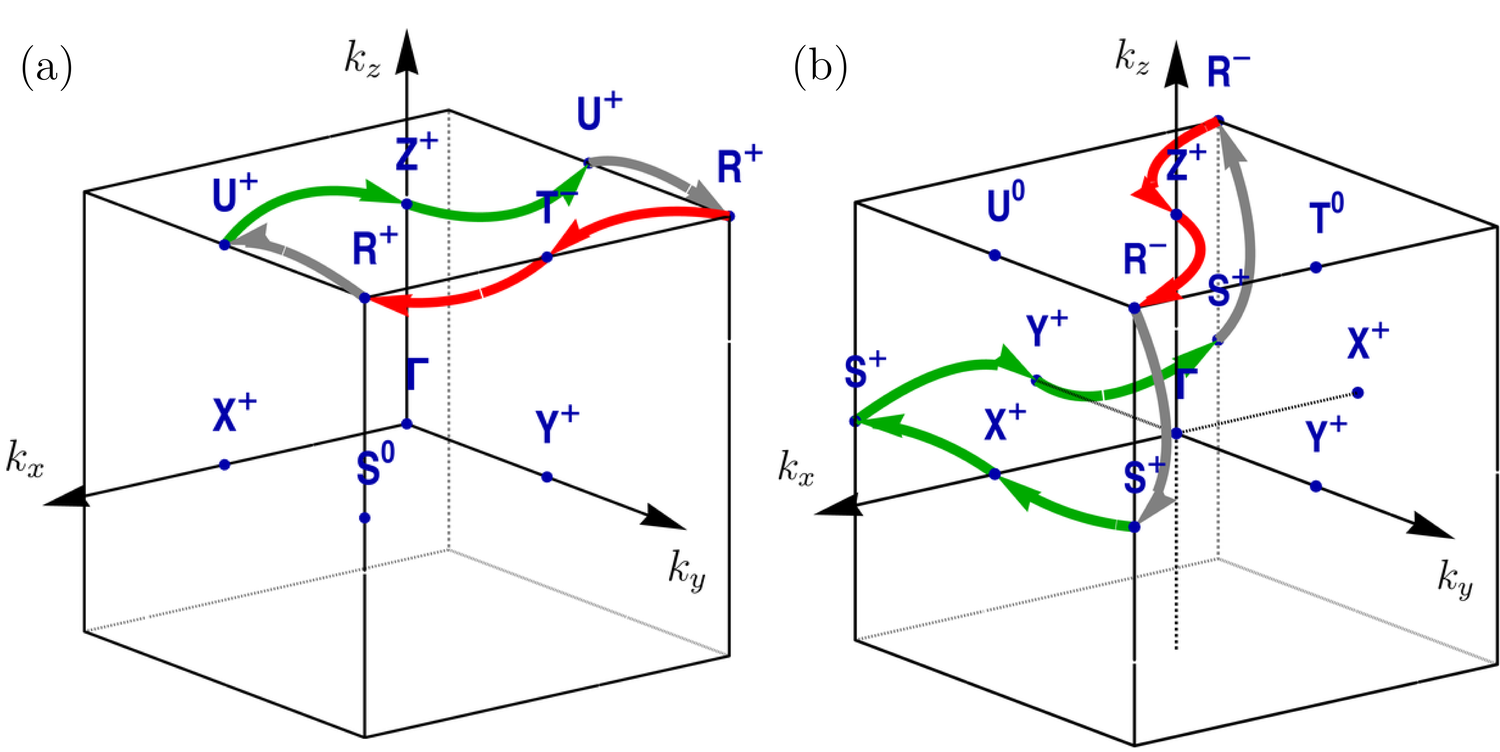}
    \caption{\justifying (a) SG 52 BZ. (b) SG 56 BZ. The symmetry-enforced $\delta_i$ products are shown in the superscripts of TRIMs labels. Points with enforced degeneracy have $0$ as the superscript. The Berry phase along the green contours is zero, along the red contours it is $\pi$. Gray lines close the loops, the total Berry phase along them is compensated.}
    \label{BerryPhaseBZ}
\end{figure}

\textcolor{black}{In this appendix we give several examples of contours with symmetry-enforced $\pi$ Berry phase for systems with negligible SOC. }
The contours contain two paths related by time-reversal symmetry. In Fig.~\ref{BerryPhaseBZ}(a) the Berry phase along contour $c_{\text{UZ}} - \overline{c_{\text{UZ}}}-(c_{\text{RT}} - \overline{c_{\text{RT}}})$ is equal to $\pi$. In the figure we show path $c_{\text{UZ}} - \overline{c_{\text{UZ}}}$ with zero Berry phase in green. Berry phase along gray lines is compensated, and along red path $-(c_{\text{RT}} - \overline{c_{\text{RT}}})$ is $\pi$. The shape of paths $c_{\text{UZ}}, \ c_{\text{RT}}$ can be arbitrary, but they must not encounter additional crossings between TRIMs.

In SG 56 and SG 138, the non-trivial invariant is enforced only along contours connecting TRIMs that can not be embedded in a 2D plane. One of the possible closed paths is shown in Fig.~\ref{BerryPhaseBZ}(b). 

We want to note that in SG 60 there are no closed contours with symmetry-enforced $\mathbb{Z}_2$ invariant (see Table~\ref{table:enforced_weakZ2_4bands}). In this group, T, U, and R points are 8-fold degenerate, and the contour must not pass these points. Z points and X, Y, and S points have different $k_z$ values (see BZ in Fig.~\ref{BerryPhaseBZ}). Therefore, a non-trivial contour includes paths connecting the Z point with one of the X, Y, and S TRIMs, and the total number of these paths quanta (that are going through only one BZ) is even. It is possible only if the total number of these quanta (with $k_z$ change) in $k_x$ and $k_y$ directions are even, too. We will refer to it as the ``$k_z$ condition".
$\pi$ Berry phase can be acquired along a path that passes S (like YSY) or starts at S (SYS). In the second case, to connect S points along a line with a trivial phase we have to add paths at the boundary of BZ (gray paths). These lines end at X or Y points. As a result, in both cases, we have to connect two X points (or Y points) separated by one BZ along a path that passes S an even number of times. This condition together with the ``$k_z$ condition" seems impossible. However, the SG 60 has non-trivial planes in the presence of a small SOC. If the coupling is enough to lift 8-fold degeneracies at T, U, and R, but is not sufficient to mix the split bands at the S point, then there are 2D planes with non-trivial $\mathbb{Z}_2$ invariant (see unstable $\mathbb{Z}_2$ invariant in Ref.~\onlinecite{PhysRevMaterials.5.124202}).

\section{Tight-binding models}

\textcolor{black}{In this appendix we present the tight-binding models for SGs 15 and 135.  }

\subsection{SG 15}\label{TBSG15}
We consider the Wyckoff position $4a$ and electron orbitals at $(0,0,0)$ and $(0,0,\tfrac{1}{2})$ sites.
In $k$ space the Hamiltonian with the spin degrees of freedom will take the form
\begin{align}
H(\textbf{k}) = \begin{pmatrix}
H_{\uparrow}(\textbf{k}) &\Delta_{\text{SOC}}\\
\Delta_{\text{SOC}}^{\dagger} & H_{\downarrow}(\textbf{k})
\end{pmatrix},
\end{align}
where $H_{\uparrow}(\textbf{k})=H_{\downarrow}(\textbf{k})=\textbf{d}\cdot\overline{\tau}$. Pauli matrices $\tau$ correspond to the orbital degree of freedom, 
\begin{align}
d_x &= 8 \cos \tfrac{k_z}{2} \left(2 \cos \tfrac{k_y}{2} \cos \tfrac{k_x+k_z}{2}+\cos \tfrac{k_z}{2}\right), \\
d_y &= 8 \sin \tfrac{k_z}{2} \left(2 \cos \tfrac{k_y}{2} \cos \tfrac{k_x+k_z}{2}+\cos \tfrac{k_z}{2}\right), \\
d_z &= -4 \sin\tfrac{k_x}{2} \sin\tfrac{k_y}{2},\\
d_0 &= 4 \cos \tfrac{k_x}{2} \cos \tfrac{k_y}{2}.
\end{align}
$\Delta_{\text{SOC}}$ is the SOC and takes the form
\begin{equation}
\Delta_{\text{SOC}} =     \left(
\begin{array}{cc}
 0 & \Delta_1 \\
 \Delta_2 & 0 \\
\end{array}
\right),
\end{equation}
where $k$-dependent functions $\Delta_1,\Delta_2$ are given by
\begin{align}
\Delta_1 &=\sqrt{2}\left(e^{-\tfrac{\pi i}{2} } \sin \tfrac{k_z}{2}+e^{\tfrac{\pi i}{2}} \cos \tfrac{k_z}{2}\right) f_{k_x k_y k_z}, \\
\Delta_2 &= \sqrt{2}\left(e^{-\tfrac{\pi i}{2} } \sin \tfrac{k_z}{2}-e^{\tfrac{\pi i}{2}} \cos \tfrac{k_z}{2}\right) f_{k_x k_y k_z}, \\
f_{k_x k_y k_z} &= \cos \tfrac{k_x+k_y+k_z}{2} +i \cos \tfrac{k_x-k_y+k_z}{2}. 
\end{align}
\subsection{SG 135}\label{TBSG135}
For SG 135 we choose the Wyckoff position 4a with coordinates $(0, 0, 0)$, $(0, 0, \tfrac{1}{2})$, $(\tfrac{1}{2}, \tfrac{1}{2}, 0)$, and $(\tfrac{1}{2}, \tfrac{1}{2},\tfrac{1}{2})$. The tight-binding model takes the form
\begin{equation}
    H(\textbf{k}) = 
        \begin{pmatrix}
            H_{\uparrow}(\textbf{k}) &\Delta_{\text{SOC}}\\
            \Delta_{\text{SOC}}^{\dagger} & H_{\downarrow}(\textbf{k})
        \end{pmatrix},
\end{equation}
where blocks correspond to spin $\uparrow$ and $\downarrow$, and are given by the matrix
\begin{align}
 H_{\uparrow}(\textbf{k})&=\left(
\begin{array}{cccc}
 h_{\uparrow11} & h_{\uparrow12} & h_{\uparrow13} & h_{\uparrow14} \\
 h_{\uparrow12}^* & h_{\uparrow22} & h_{\uparrow23} & h_{\uparrow13}\\
 h_{\uparrow13}^* & h_{\uparrow23}^*& h_{\uparrow11} &h_{\uparrow12} \\
 h_{\uparrow14}^* & h_{\uparrow13}^* & h_{\uparrow12}^* & h_{\uparrow22} \\
\end{array}
\right). 
\end{align}
The matrix elements are $k$-dependent functions
\begin{align*}
h_{\uparrow11} &= h_{\downarrow11} = 2\cos k_x t_7+2\cos k_y t_6+2\cos k_z t_8+t_1,\\
h_{\uparrow22} &=  h_{\downarrow22} = 2\cos k_x t_6+2\cos k_y t_7+2\cos k_z t_8+t_1,\\
h_{\uparrow12} &= h_{\downarrow12} = \left(1+e^{-i k_z}\right) t_2,\\
h_{\uparrow13} &=\left(1+e^{i k_x}\right) \left(1+e^{i k_y}\right) t_{3}^* e^{-i (k_x+k_y)},\\
h_{\uparrow14} &=\left(1+e^{i k_z}\right) e^{-i (k_x+k_y+k_z)} \\
 &\cross \left[t_{4}^* \left(1+e^{i (k_x+k_y)}\right)+t_{5}^* \left(e^{i k_x}+e^{i k_y}\right)\right],\\
h_{\uparrow23} &=\left(1+e^{i k_z}\right) e^{-i (k_x+k_y)} \\
 &\cross \left[t_{4}^* \left(e^{i k_x}+e^{i k_y}\right)+t_{5}^* \left(1+e^{i (k_x+k_y)}\right)\right],\\
h_{\downarrow13} &=\left(1+e^{i k_x}\right) \left(1+e^{i k_y}\right) t_{3} e^{-i (k_x+k_y)},\\
h_{\downarrow14} &=\left(1+e^{i k_z}\right) e^{-i (k_x+k_y+k_z)} \\
 &\cross \left[t_{4} \left(1+e^{i (k_x+k_y)}\right)+t_{5} \left(e^{i k_x}+e^{i k_y}\right)\right],\\
  \end{align*}
 \begin{align*}
h_{\downarrow23} &=\left(1+e^{i k_z}\right) e^{-i (k_x+k_y)} \\
 &\cross \left[t_{4} \left(e^{i k_x}+e^{i k_y}\right)+t_{5} \left(1+e^{i (k_x+k_y)}\right)\right],\\
\end{align*}
with parameters given by
\begin{align*}
t_1 &= 0, t_2 = -0.024, t_3 = 0.594 - 0.366 i, \\
t_4 &= 0.372 - 0.368 i,
t_5 = -0.395 + 0.895 i, \\
t_6 &= -1.414, 
t_7 = 1.962, 
t_8 =1.966.
\end{align*}
The SOC is described by the matrix
\begin{align}
\Delta_{\text{SOC}} = \left(
\begin{array}{cccc}
 0 & 0 & 0 & \Delta_1 \\
 0 & 0 &\Delta_2 & 0 \\
 0 & \Delta_3 & 0 & 0 \\
 \Delta_4 & 0 & 0 & 0 \\
\end{array}
\right)
\end{align}
with $k$-dependent elements
\begin{align*}
\Delta_1 &= \left(-1+e^{i k_z}\right) e^{-i (k_x+k_y+k_z)} \\
 &\cross\left[t_{9} \left(-1+e^{i (k_x+k_y)}\right)+t_{10} \left(e^{i k_x}-e^{i k_y}\right)\right],\\
\Delta_2 &=  i\left(-1+e^{i k_z}\right) e^{-i (k_x+k_y)} \\
 &\cross\left[t_{9} \left(e^{i k_x}-e^{i k_y}\right)-t_{10} \left(-1+e^{i (k_x+k_y)}\right))\right],\\
\Delta_3 &= e^{-i k_z} \left(-1+e^{i k_z}\right) \\
 &\cross\left[t_{9}^* \left(e^{i k_x}-e^{i k_y}\right)+t_{10}^* \left(-1+e^{i (k_x+k_y)}\right)\right],\\
\Delta_4 &= -i \left(-1+e^{i k_z}\right) \\
 &\cross\left[t_{9}^* \left(-1+e^{i (k_x+k_y)}\right)-t_{10}^* \left(e^{i k_x}-e^{i k_y}\right)\right].\\
\end{align*}
The parameters for our model are given by
\begin{align*}
t_{9} = -0.688 - 0.688 i,
t_{10} = -0.098 + 0.098 i.
\end{align*}
\section{\textcolor{black}{Material examples}}\label{candidates}

\textcolor{black}{In this appendix we present the band structures for several material candidates. In each figure, we highlighted topological gaps, which are every second gap in the band structures. We also show the $\delta_i$ [see Eq.~(\ref{Z2_invariant})] values for TRIMs in the topological planes.
}

\textcolor{black}{To compute the band structures, we used the Quantum ESPRESSO software~\cite{Giannozzi_2009,Giannozzi_2017}. We used PAW~\cite{PhysRevB.50.17953} and ultrasoft~\cite{PhysRevB.41.7892} pseudopotentials. The pseudopotentials are listed in the footnote~\cite{TaPpseudo}.
We obtained the $\delta_i$ values with the irrep software~\cite{IRAOLA2022108226,Elcoro:ks5574}. The ICSD codes for the materials are the following 24327 (P$_2$Pt$_5$), 408695 (CsHg$_2$), 108656 (TaP), 615698 (BW), 108147 (CaPt$_2$) and 649849 (YPt$_2$).}
\begin{figure}[htbp]
    \centering
\includegraphics[scale=0.22]{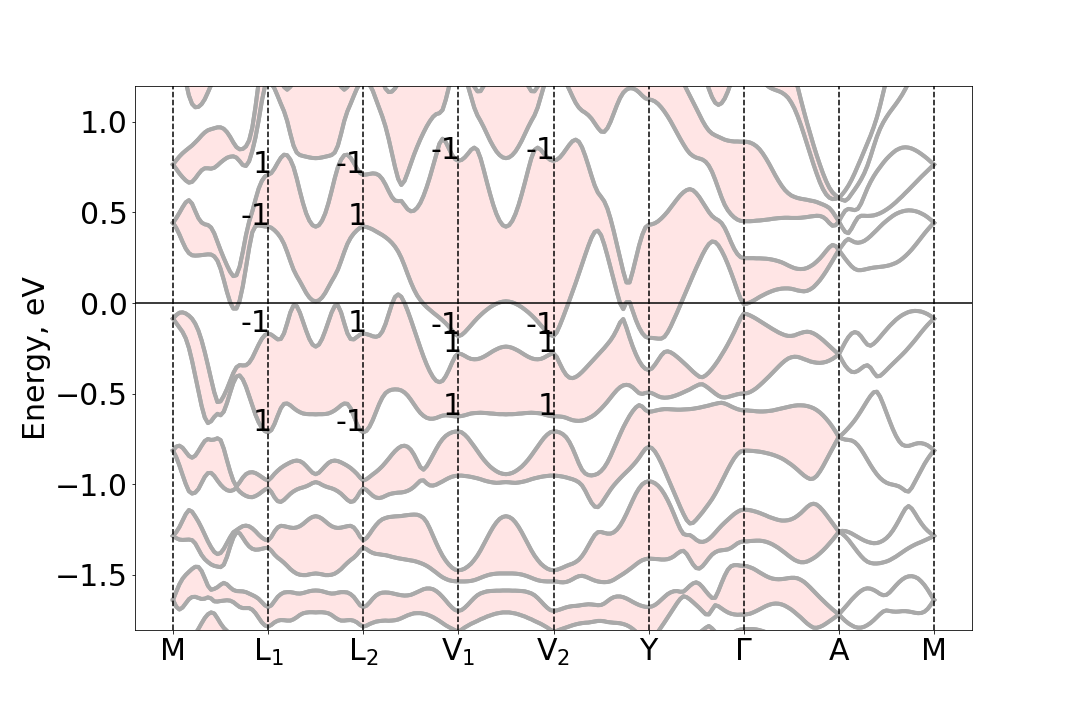}
\caption{\justifying \textcolor{black}{Band structure of P$_2$Pt$_5$ in SG 15. Numbers correspond to the values $\delta_{L}$ and $\delta_{V}$. The coordinates of the TRIMs in the primitive basis L$_1^g(\pi, 0, \pi)$, L$_2^g(0, \pi, -\pi)$, V$_1^g(\pi, 0, 0)$ and V$_2^g(0, \pi, 0)$.
}}
\label{P2Pt5_bands}
\end{figure}
\begin{figure}[htbp]
    \centering
\includegraphics[scale=0.22]{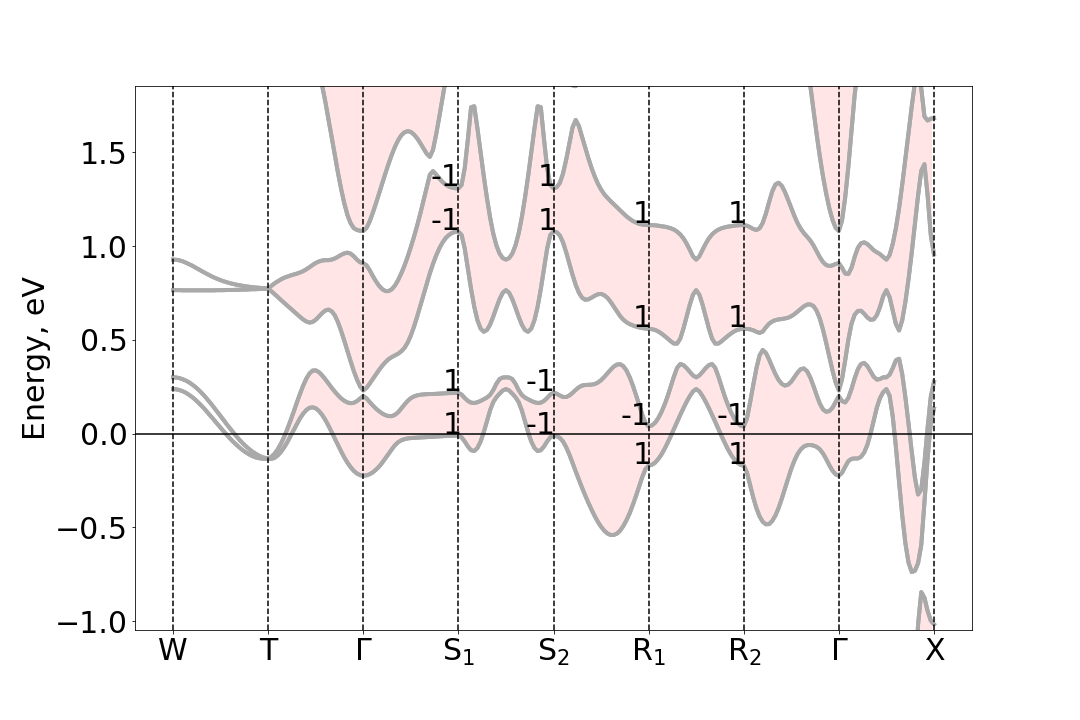}
\caption{\justifying \textcolor{black}{Band structure of CsHg$_2$ in SG 74. Numbers correspond to the values $\delta_{R}$ and $\delta_{S}$. The coordinates of the TRIMs in the primitive basis S$_1^g(\pi, 0, 0)$, S$_2^g(0, \pi, \pi)$, R$_1(0, \pi, 0)$ and R$_2(\pi, 0, \pi)$. 
}}
\label{CsHg2_bands}
\end{figure}
\begin{figure}[htbp]
    \centering
\includegraphics[scale=0.22]{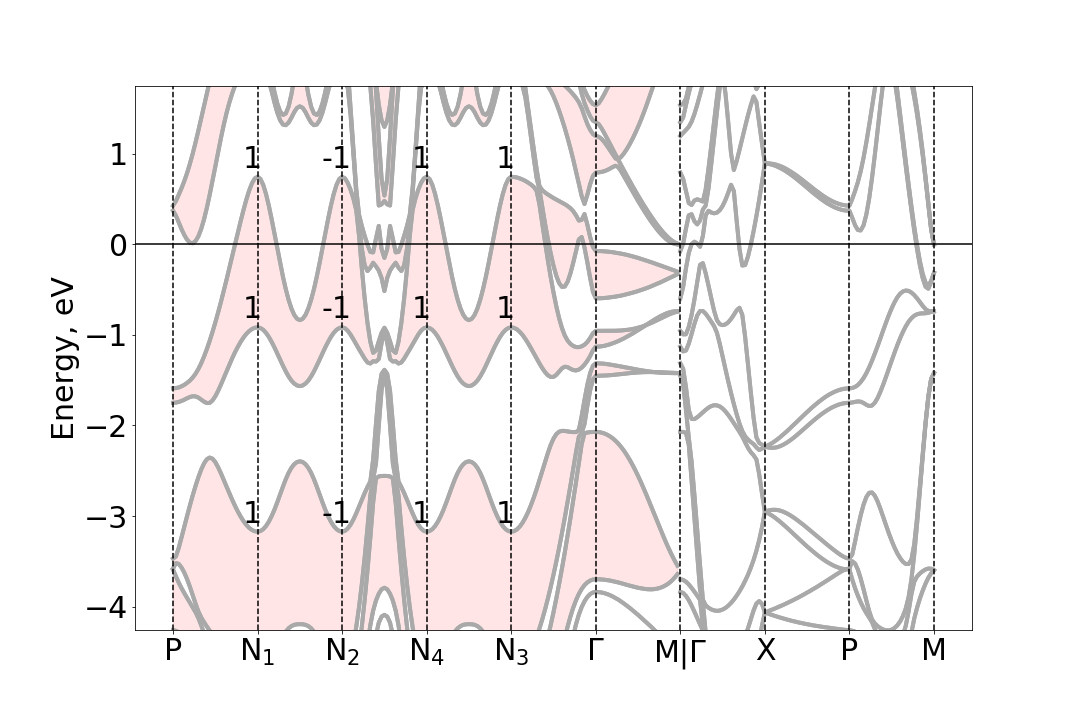}
\caption{\justifying \textcolor{black}{Band structure of TaP in SG 141. Numbers correspond to the values $\delta_{N}$. The coordinates of the TRIMs in the primitive basis N$_1^g(0, \pi, 0)$, N$_2^g(\pi, 0, 0)$, $N_3^g(\pi, 0, -\pi)$ and N$_4^g(0, \pi, -\pi)$. 
}}
\label{TaP_bands}
\end{figure}
\begin{figure}[htbp]
    \centering
\includegraphics[scale=0.22]{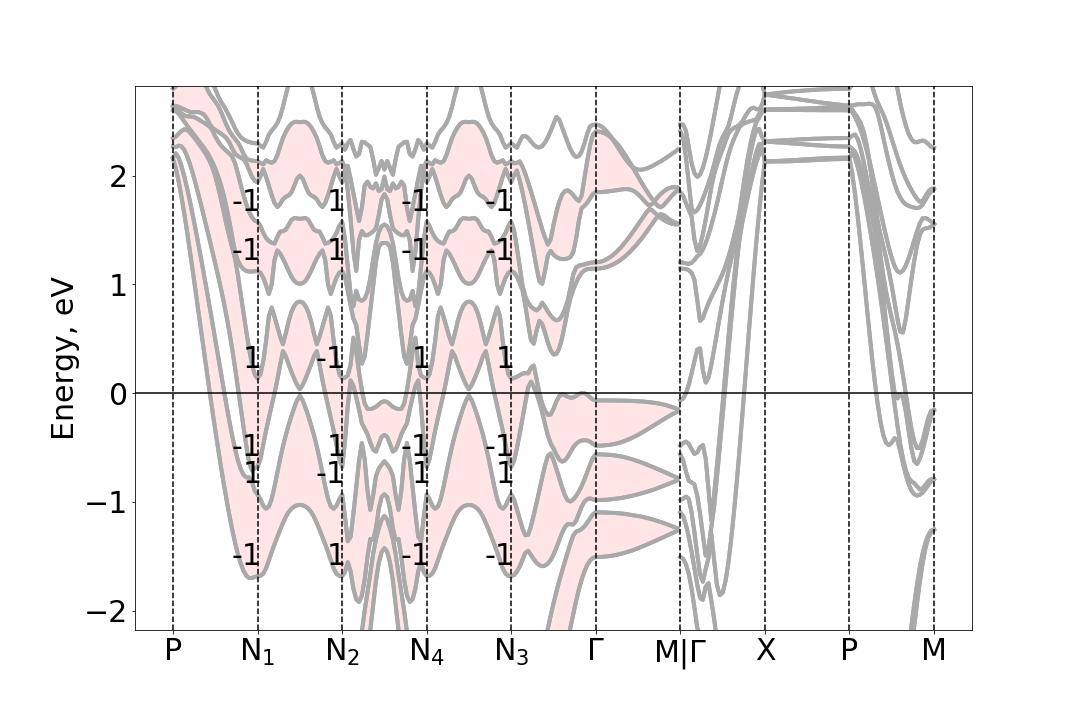}
\caption{\justifying \textcolor{black}{Band structure of BW in SG 141. Numbers correspond to the values $\delta_{N}$. The coordinates of the TRIMs in the primitive basis N$_1^g(0, \pi, 0)$, N$_2^g(\pi, 0, 0)$, $N_3^g(\pi, 0, -\pi)$ and N$_4^g(0, \pi, -\pi)$. 
}}
\label{BW_bands}
\end{figure}
\begin{figure}[htbp]
    \centering
\includegraphics[scale=0.22]{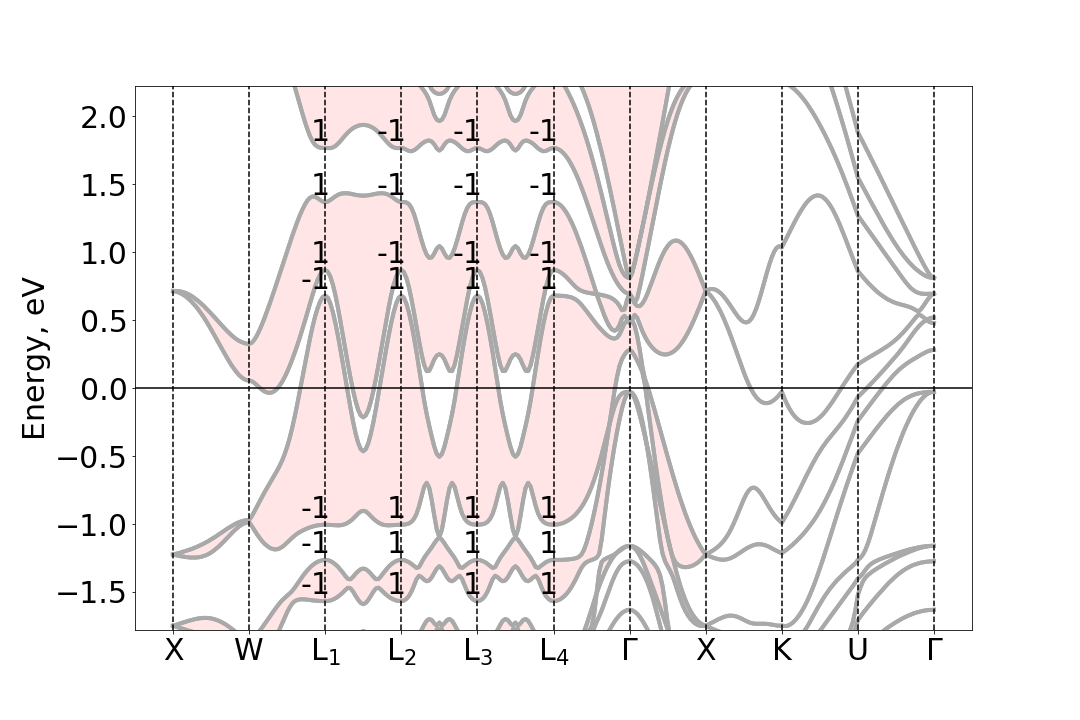}
\caption{\justifying \textcolor{black}{Band structure of CaPt$_2$ in SG 227. Numbers correspond to the values $\delta_{L}$. The coordinates of the TRIMs in the primitive basis L$_1^1(\pi, \pi, \pi)$, L$_2^g(0, 0, \pi)$, L$_3^g(\pi, 0, 0)$ and L$_4^g(0, \pi, 0)$.
}}
\label{CaPt2_bands}
\end{figure}
\begin{figure}[htbp]
    \centering
\includegraphics[scale=0.22]{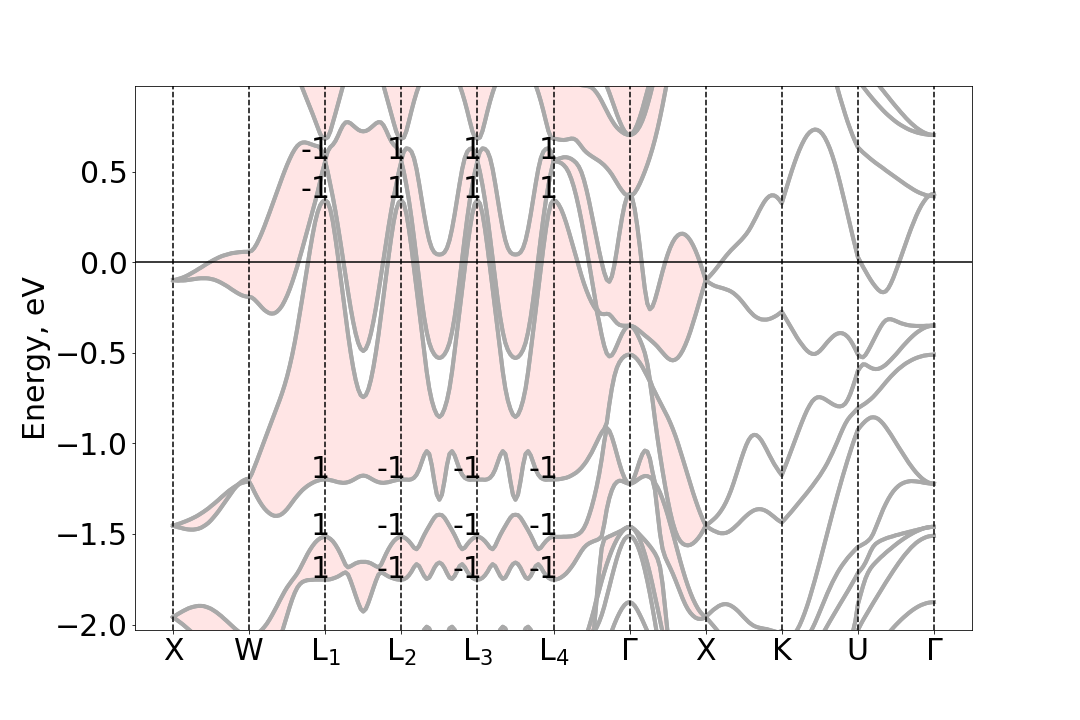}
\caption{\justifying \textcolor{black}{Band structure of YPt$_2$ in SG 227. Numbers correspond to the values $\delta_{L}$. The coordinates of the TRIMs in the primitive basis L$_1^1(\pi, \pi, \pi)$, L$_2^g(0, 0, \pi)$, L$_3^g(\pi, 0, 0)$ and L$_4^g(0, \pi, 0)$.
}}
\label{Pt2Y_bands}
\end{figure}

\bibliography{ref_v2.bib}

\end{document}